\documentclass{article}

\usepackage[english]{babel}
\usepackage{amssymb} 
\usepackage[letterpaper,top=2cm,bottom=2cm,left=3cm,right=3cm,marginparwidth=1.75cm]{geometry}
\usepackage[T2A,T1]{fontenc}
\usepackage{tabularx}
\usepackage{graphicx}
\usepackage{lineno,hyperref}
\usepackage{bm}
\usepackage{color}
\usepackage{tcolorbox}
\usepackage{amsmath}
\usepackage{epstopdf}
\usepackage{adjustbox}
\usepackage{newfloat}
\usepackage{epsfig}
\usepackage{array}
\usepackage{amsmath}
\usepackage{graphicx}
\usepackage{hyperref}
\hypersetup{colorlinks,allcolors=black}

\title{Modelling the Impact of Organic Molecules and Phosphate Ions on Biosilica Pattern Formation in Diatoms}

\author{Svetlana Petrenko  \and Karen M. Page}

\date{Department of Mathematics
, University College London
, Gower Street, London WC1E 6BT
, UK}
\begin{document}
\maketitle

\begin{abstract}
The rapid and complex patterning of biosilica in diatom frustules is of great interest in nanotechnology, although it remains incompletely understood. Specific organic molecules, including long-chain polyamines, silaffins, and silacidins are essential in this process. The molecular structure of the synthesized polyamines significantly affects the quantity, size, and shape of silica precipitates. Experimental findings show that silica precipitation occurs at specific phosphate ion concentrations.
We focus on the hypothesis that pattern formation in diatom valve structures is driven by phase separation of species-specific organic molecules. The resulting organic structures serve as templates for silica precipitation. We investigate the role of phosphate ions in self-assembly of organic molecules and analyze how the reaction between them affects the morphology of the organic template. Using mathematical and computational techniques, we gain an understanding of the range of patterns that can arise in a phase-separating system. By varying the degree of dissociation and the initial concentrations of reacting components we demonstrate that the resulting geometric features are highly dependent on these factors. This approach provides insights into the parameters controlling patterning. Additionally, we consider the effects of prepatterns, mimicking silica ribs that preexist the pores, on the final patterns.
\end{abstract}

\maketitle


\section{Introduction}

\begin{figure*}[h]
	\begin{center} \
	\begin{minipage}[h]{1\linewidth} 
		{\includegraphics[scale=0.5]{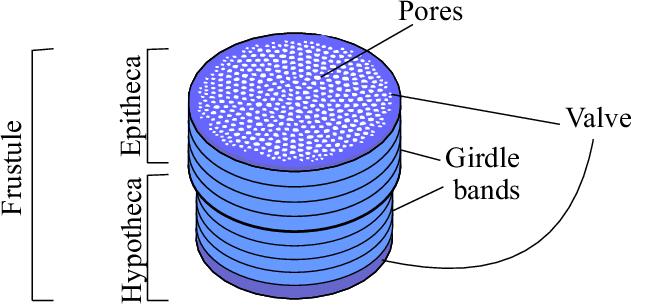}} 
	\end{minipage}
	\caption{Schematic structure of the diatom frustule \label{diagram}}\end{center}
\end{figure*}

Intricately patterned biosilica  in diatom frustules is of great interest to science and nanotechnology, however, its mode of formation remains unclear. Diatoms are unicellular algae that have been extensively studied for their ability to form biosilica. They are found in aquatic environments that have sufficient sunlight and minerals. Their silica cell walls consist of two thecae composed of intricately shaped valves and girdle bands (see Fig.~\ref{diagram}) \cite{round1990diatoms}, perform various functions. They have the mechanical strength to resist shear stress and provide protection against predators \cite{de2009functional,hamm2003architecture,garcia2010bioinspired}. The porous structure of cell walls facilitates the filtration and sorting of nutrients \cite{hale2001functional}. The valves themselves are formed by a system of silica ribs, costae, connected by tiny silica bridges, resulting in self-supporting structures that use minimum amounts of silica. Furthermore, in the areas between the ribs the valves exhibit distinctive patterns characterized by self-similar pores.

Due to their ecological importance diatoms are often called "the lungs of the Earth". These microorganisms  play a significant role in the global carbon cycle \cite{milligan2002proton,treguer2018influence} and in climate regulation. They are responsible for approximately 40\% of primary production and 20-30\% of the oxygen we breathe \cite{alverson2014air,benoiston2017evolution}, emphasizing their key role in the maintenance of human existence.

However, diatoms have even greater potential due to their exceptional optical and electrical properties. Their meticulously structured porous silica valves are optimized for light focusing and harvesting. The pore pattern within the valve functions as an array of microlenses \cite{de2007lensless,de2008light}, diffracting incoming photons in a manner analogous to that of a photonic crystal \cite{joannopoulos1997photonic}. This unique characteristic can be used in a wide range of optical applications, ranging from advanced imaging systems to sensor technologies \cite{rea2019recent}.

In addition to their optical, electrical, structural and mechanical properties, diatom silica valves are sustainable, cost-effective \cite{wang2017prospects} and eco-friendly materials \cite{watanabe1996photosynthetic}. The renewable nature of diatoms and their biomineralization processes represent a compelling paradigm for sustainable materials production. Diatom-based materials have potential applications as biosensors \cite{losic2009diatomaceous,rea2019recent}, drug delivery systems \cite{zobi2022diatom,phogat2021diatom,delasoie2019natural} and support materials for advanced technological devices \cite{rabiee2021diatoms}. Their distinctive properties, including photoluminescence \cite{jeffryes2008electroluminescence}, are of significant value for a wide range of applications.

Unlocking the secrets of diatom morphogenesis is important in addressing climate change, developing renewable energy, creating advanced materials, enhancing biomedicine, etc. Equipped with this knowledge, a greener, healthier and more sustainable future can be envisioned.

The formation of diatom cell walls occurs within specialized compartments known as silica deposition vesicles (SDVs). Silicic acid, a soluble form of silica, is taken up by silicon transporters into the SDV where it is precipitated. Diatoms use sophisticated mechanisms to prevent uncontrolled precipitation and to maintain hight intracellular concentrations of silicic acid. The proposal that organic components interact with silicic acid to stabilize it within silica transport vesicles provides insight into the formation of silica spheres, which promote cell wall growth \cite{schmid1979wall}.

Recent advances in diatom research have elucidated the molecular mechanisms underlying pore pattern morphogenesis in the diatom silica \cite{heintze2022molecular}. This intricate process begins during the early stages of costa formation and centers around organic droplets within SDVs. During this stage, the space between the silica ribs does not contain precipitated silica, but organic droplets are formed through a liquid-liquid phase separation process. These droplets contain long chain polyamines (LCPAs), are positively charged and arrange themselves along the negatively charged silica rib edges, maintaining an equal spacing through electric repulsion. As they take position, they provide the basis for the formation of the cribrum matrix, which covers the entire intervening space of the silica ribs, except where the organic droplets are located.

The silica precursor enters the space between the ribs, however its exact composition remains uncertain. The silica nanoparticles found in SDVs indicate their potential role as feedstock for silica growth in the SDV. At this stage, the organic droplets act as templates, preventing the infiltration of the silica precursor and guiding the formation of pores. The unique precursor pores exhibit wave-like peaks and anvil-shaped features, which emerge as a result of organic droplets arresting the silica deposition. Silicification starts at the edges of the ribs and advances toward the opposite silica ribs. The matrix of the cribrum plate is silicified continuously.

Chemical analysis of diatom biosilica revealed a diverse set of organic components, including carbohydrates, lipids, amino acids, and proteins, specifically $\alpha 1$ -frustulin, which is related to the biosilica matrix, \cite{volcani1981cell}. The formation of silica structures is influenced by species-specific organic molecules, such as LCPAs, silaffins, and silacidins, which play a key role in templating of silica deposition \cite{pawolski2018reconstituting,kroger2010diatom}.

The molecular structure of the synthesized polyamines significantly affects the amount, size, and morphology of silica precipitates \cite{bernecker2010tailored}. In addition, phosphate ions are essential for silica precipitation \cite{sumper2006learning,bernecker2010tailored,wieneke2011silica}. The pH threshold required for silica precipitation was shown to depend on phosphate concentration \cite{sumper2006learning}. Higher concentrations of phosphate led to the formation of larger spherical aggregates of organic molecules,  which serve as templates for silica formation upon addition of silicic acid molecules to the solution \cite{sumper2006learning}. The size of polyamine aggregates can be controlled by regulating the pH level at a fixed polyamine/phosphate ratio \cite{sumper2006learning}. Furthermore, an experimental study on silaffins \cite{kroger2002self} showed that, unlike natSil-1A, silaffin-1A could not induce silica precipitation in the absence of phosphate. Considering these experimental results, it is evident that the interaction between organic molecules and phosphate ions significantly influences the phase separation of the organic molecules.

Considerable efforts have been made to elucidate the complex chemical and biological processes involved in the silica morphogenesis in diatom valves. Despite these advances, theoretical modelling of diatom valve morphogenesis remains quite underexplored.

One pioneering theory proposed diffusion-limited aggregation (DLA) of silica nanoparticles  at the leading edges of the SDV as a model for silica rib formation \cite{gordon1994chemical}. This model describes the random fluctuation of silica particles in a fluid environment, leading to the formation of dendritic structures. This suggests that a nucleation structure in the center of the SDV initiates the formation of silica ribs via aggregation and sintering processes.

The laterally dependent stochastic aggregation (LSDA) model, a probabilistic model, which focus on silica precipitation and dissolution rates at solid/fluid interfaces, provided valuable insights into porous structures \cite{willis2013simple}. Pore formation is promoted by negative lateral feedback in the solid regions, whereas the fluid regions encouraged silicification. Potential chemical reagents inside the SDVs have been suggested to control this feedback.



The hypothesis of organic molecules guiding silica pattern formation within SDVs introduced the concept of liquid-liquid phase separation as a mechanism for pattern template formation \cite{sumper2002phase,lenoci2006self,bobeth2020continuum}. The models suggest that these processes create organic phases with silica precipitation activity and aqueous phases containing silica precursors. Silica precipitates around the organic droplets, forming pore cavities. The work carried out in \cite{lenoci2006self} yielded numerous two-dimensional patterns that closely resembled the observed valve structures, where phase separation was induced by an additional local field arising from the pre-existing silica costae in the base layer. Another more recent model \cite{bobeth2020continuum}, which couples phase separation with a chemical reaction, aimed to study the role of phosphate ions in the self-assembly of LCPAs. While two- and three-dimensional simulations of pattern formation have been presented, the influence of model parameters on the morphology of the organic template remains unclear.

Building upon these previous works, our study seeks to elucidate the roles of organic molecules and phosphate ions in the formation of diatom valve silica. Expanding on the framework of \cite{bobeth2020continuum} our aim is to demonstrate how the model parameters influence pattern formation.

This paper is structured as follows. First, to study the role of phosphate ions in the aggregation process, we introduce a comprehensive mathematical model of the phase separation of organic molecules coupled with a chemical reaction. Subsequently, we examine the free energy associated with this process using the Flory-Huggins model. Following this, we conduct a linear stability analysis of the model equations to obtain the necessary conditions for pattern formation. The next section is devoted to numerical simulations to study the various stationary patterns obtained. Furthermore, we analyze the influence of the confined geometry, model parameters, and prepatterning field on the resulting patterns.

\section{Model}

Our research aims to analyze the aggregation of organic molecules. To understand the influence of phosphate ions on this phenomenon, we studied a simplified and generic chemical reaction:
\begin{gather} \label{reaction}
    A+B \overset{\beta}{\underset{\alpha}{\leftrightarrows}} C, 
\end{gather}
where $A$ represents the organic molecules that cannot undergo phase separation.
We assume that when phosphate ions (component $B$) bind to these organic molecules, a new organic component (component $C$) is formed with a higher level of phosphorylation, see \cite{bobeth2020continuum}. This increased phosphorylation decreases the charge of component $C$, making it more likely to undergo phase separation.
The dynamics of system \eqref{reaction} is determined by the continuity equation:
\begin{gather} 
    \cfrac{\partial \mathbf{c}}{\partial t} = -\nabla \cdot \mathbf{j} + \mathbf{r},
\end{gather}
where $\mathbf{c}=(c_A, c_B, c_C)$, with $c_A$, $c_B$ and $c_C$ as concentrations of the chemical components $A$, $B$ and $C$, respectively; $\mathbf{j}$ denotes spatial fluxes, which are driven by diffusive processes for species $A$ and $B$; $\mathbf{r}$ denotes the source term related to chemical reactions that convert new organic component into a mixture of solvent and organic molecules able to phase separate. The chemical reactions are local, described by rate laws (depending on the composition).

The source term $\mathbf{r}$ is defined by \eqref{reaction}.  The reaction velocity of components $A$ and $B$ is described by coefficient $\alpha$, and the reverse reaction (dissociation of component $C$) is described by coefficient $\beta$. Therefore, from \eqref{reaction} it follows that:
\begin{gather}
\mathbf{r}=(-\alpha c_A c_B+\beta c_C, -\alpha c_A c_B + \beta c_C, \alpha c_A c_B - \beta c_C).    \nonumber
\end{gather}

For organic molecules and phosphate ions the diffusion flux $\mathbf{j}$ is given by Fick’s law: $\mathbf{j}=-D\nabla \mathbf{c}$, where $D$ is diffusion coefficient. Therefore, the evolution of the concentrations of components $A$ and $B$ is described by the following reaction-diffusion equations:
\begin{gather}\label{c_A}
\cfrac{\partial c_A}{\partial t}= \nabla \cdot D_A \nabla c_A -\alpha c_A c_B + \beta c_C \\
\cfrac{\partial c_B}{\partial t}= \nabla \cdot D_B \nabla c_B -\alpha c_A c_B + \beta c_C \label{c_B},
\end{gather}
where $D_A$ and $D_B$ are diffusion coefficients, which are typically dependent on the concentrations $c_A$, $c_B$ and $c_C$ and the charge state of the polymer. However, because specific information on the studied polyamine-phosphate ion-water system is not available, we assume in this analysis that these kinetic coefficients are constant.

To describe the phase separation process of component $C$ the Cahn-Hilliard model was used. The volume fraction $\phi$ of component $C$ is related to its concentration by $\phi=\Omega_c c_C$ with $ \Omega_c$ is the molecular volume. This model is particularly useful for understanding the thermodynamic behaviour of the system, and it allows us to study the effects of different parameters on the phase separation process. The thermodynamic state of phase separating component is governed by the free energy:
\begin{gather} \label{free_energy}
    F[\phi]=\int \left( f(\phi) +\cfrac{\kappa}{2} (\nabla \phi)^2\right) dV,
\end{gather}
where $\phi(\mathbf{r})$ is the volume fraction of component $C$, $f(\phi)$ is the free energy density of mixing and $\kappa$ is the gradient energy coefficient (assumed to be constant).

For component $C$, the non-local diffusive fluxes $j$ are driven by a gradient in the chemical potential. The linear non-equilibrium thermodynamics implies $j=-M(\phi)\nabla\mu$, where $M$ is the positive mobility (for simplicity, it is approximated as a constant) and $\mu$ is the chemical potential. Hence we obtain
\begin{gather}
    \cfrac{\partial \phi}{\partial t} = \nabla \cdot \left( M \nabla \mu \right) + \alpha c_A c_B - \beta \phi.
\end{gather}

Together with the constraint $\cfrac{1}{V} \int dV \phi(\mathbf{r})=\overline{\phi}$ and since $\mu=\delta F/ \delta \phi$, we obtain the evolution equation for the volume fraction $\phi(\mathbf{r},t)$:
\begin{gather}\label{phi}
  \cfrac{\partial \phi}{\partial t}=\nabla \cdot M \nabla \left( f'(\phi) - \kappa \nabla^2\phi \right) +\alpha c_A c_B - \beta \phi  .
\end{gather}

We employ the Flory-Huggins free energy density:
\begin{gather} \label{g_FH}
  f(\phi)=kT \left[ \cfrac{1}{N}\phi \ln \phi + (1-\phi) \ln (1-\phi) +\chi \phi (1-\phi) \right],  
\end{gather}
where 
$k$ is the Boltzmann constant, $T$ is the absolute temperature, $N$ 
is the length of polymer in units of the solvent size and $\chi$ is the Flory-Huggins interaction parameter. 

Spinodal decomposition is used in this study. This refers to a process in which a single thermodynamic phase naturally splits into two phases without the need for nucleation. This separation occurs in the absence of a thermodynamic barrier for phase separation. Spinodal decomposition is observed when the second derivative of the free energy density is negative, denoted as $f''(\phi)<0$.

To make the equations dimensionless, we introduce the length unit $l=\sqrt{\kappa/(kT)}$, the time unit $\widehat{t}=t/\tau$ with $\tau=l^2/(M(kT))$, $\widehat{\nabla}=l \nabla$, and $\widehat{f}=f/(kT)$. This allows us to express the reaction-diffusion equations \eqref{c_A} and \eqref{c_B} for $c_A$, $c_B$ multiplyed by $\Omega_C$ and the evolution equation \eqref{phi} for the volume fraction $\phi$ in a dimensionless form:

\begin{gather} \label{ca_eq}
 \cfrac{\partial \widehat{c}_A }{\partial \widehat{t}}= \widehat{\nabla}  \cdot \widehat{D}_A \widehat{\nabla} \widehat{c}_A - \widehat{\alpha} \widehat{c}_A \widehat{c}_B + \widehat{\beta} \phi  \\ \label{cb_eq}
 \cfrac{\partial \widehat{c}_B }{\partial \widehat{t}}= \widehat{\nabla}  \cdot \widehat{D}_B \widehat{\nabla} \widehat{c}_B - \widehat{\alpha} \widehat{c}_A \widehat{c}_B + \widehat{\beta} \phi \\
 \cfrac{\partial \phi}{\partial \widehat{t}} = \widehat{\nabla}^2 \left( \widehat{f}'(\phi) - \widehat{\nabla}^2 \phi \right) +  \widehat{\alpha} \widehat{c}_A \widehat{c}_B - \widehat{\beta} \phi \label{phi_d},
\end{gather}
where $\widehat{c}_A=\Omega_C c_A$, $\widehat{c}_B=\Omega_C c_B$ are the concentrations $c_A$ and $c_B$ scaled by the molecular volume of component $C$; $\widehat{D}_A=D_A \tau/l^2$, $\widehat{D}_B =D_B \tau/l^2$, $\widehat{\alpha}= \alpha \tau / \Omega_C$ and $\widehat{\beta}=\beta\tau$. This model was used in \cite{bobeth2020continuum}.

 We numerically solve the system of nonlinear partial differential equations \eqref{ca_eq}-\eqref{phi_d}, which describe the coupled evolution of the three components. The finite element software COMSOL Multiphysics was used to conduct simulations in confined geometry. Our phase separation model is distinguished by six model parameters: $\chi$, $N$, $D_A$, $D_B$, $\alpha$, and $\beta$, together with the initial concentrations $c_{A,0}$ and $c_{B,0}$. Given the significant number of parameters involved, our initial analysis focuses on determining the conditions under which a uniform system becomes unstable, leading to spinodal decomposition.

\section{Flory-Huggins model}

Before analyzing the linear stability, we thoroughly examine the free energy using the Flory-Huggins model.

The Flory-Huggins model is a fundamental theoretical framework that elucidates the intricate behaviour of liquid-liquid phase separation, which emerges from the interplay between entropy and interaction energy. In this model, we examine the blending of a polymer species of length $N$ with a homogeneous solvent.
The Flory-Huggins model employs an effective lattice site contact energy between the polymer and solvent, denoted as $\chi\equiv\cfrac{z}{2}\cfrac{2\epsilon_{ps}-\epsilon_{ss}-\epsilon_{pp}}{kT}$ in which $z$ is a coordination constant, and $\epsilon_{ps}$, $\epsilon_{ss}$ and $\epsilon_{pp}$ are bare polymer-solvent, solvent-solvent and polymer-polymer contact energies. The free energy density of the Flory-Huggins model is given by \eqref{g_FH}.

On examination of the right-hand side of \eqref{g_FH}, we discern that the first two terms signify the entropic free energy of mixing, whereas the third term represents the effective contact energy. The spinodal concentration defines the boundary between the locally stable and unstable regions and can be precisely determined.

The free energy density becomes locally unstable when $f''(\phi)\leq 0$, and the spinodal boundary $\phi^{*}$ is defined at the transition point $f''(\phi^{*})=0$
\begin{gather} \label{g_sec_d}
   f''(\phi)=\cfrac{1}{N\phi}+\cfrac{1}{1-\phi}-2\chi.
\end{gather}
Solving for this condition, we obtain dense $(\phi^{s}_1)$ and dilute phase $(\phi^{s}_2)$ spinodal concentrations 
\begin{gather} \label{fi12}
   \phi^{s}_{1,2}=\cfrac{1}{4\chi N}\left(N(2\chi-1)+1 \pm \sqrt{(N(2\chi -1) +1)^2-8\chi N}\right) .
\end{gather}

The critical point of liquid-liquid phase separation occurs when the dense and dilute phases coincide, corresponding to a critical interaction strength $\chi^*$ and concentration $\phi^*$:
\begin{gather}
\chi^*=\cfrac{1}{2}\left(1+\cfrac{1}{N}+\cfrac{2}{\sqrt{N}}\right), \qquad \phi^* =\cfrac{\sqrt{N}+1}{N+1+2\sqrt{N}}.    \nonumber
\end{gather}

Near the critical point with $\delta \chi \equiv \chi - \chi^*\approx 0$, the spinodal concentrations have the approximate form
\begin{gather}
\phi^s_{1,2}=\phi^* \pm \sqrt{\cfrac{\delta\chi}{2(\chi^*)^2\sqrt{N}}} +O(\delta\chi).    \nonumber
\end{gather}

 Fig.~\ref{g(f)} illustrates how the Flory-Huggins interaction parameter affects spinodal decomposition for a given value of $N$.

For simplicity, we assume that the reaction and diffusion processes are faster than phase separation. Therefore, we begin with the following initial conditions:
\begin{gather}
   \left.c_A\right|_{t=0}=c_{A,0}, \qquad \left.c_B\right|_{t=0}=c_{B,0}, \qquad \left.\phi\right|_{t=0}=0. \nonumber
\end{gather}

Following this, a rapid establishment of reaction equilibrium takes place:
\begin{gather}
    \alpha c_{A,e} c_{B,e}=\beta \phi_e. \nonumber
\end{gather}

Considering the conservation law of the particles, we derive:
\begin{gather}
 c_{A,0}=c_{A,e}+\phi_e, \qquad c_{B,0}=c_{B,e}+\phi_e. \nonumber  
\end{gather}

By considering both the initial and equilibrium conditions, we can deduce the equilibrium concentration of component $C$ as:
\begin{gather} \label{f_eq}
    \phi_e=\cfrac{1}{2}\left( c_{A,0} +c_{B,0}+\cfrac{\beta}{\alpha} - \sqrt{\left( c_{A,0} +c_{B,0} + \cfrac{\beta}{\alpha} \right)^2-4 c_{A,0} c_{B,0}} \right).
\end{gather}

Keeping $\chi$ and $N$ fixed and substituting $\phi$ with $\phi_e$ in \eqref{g_sec_d}, we can examine $f''(\phi_e(c_{A,0},c_{B,0},\alpha,\beta))$. This allows us to determine the range of values for $c_{A,0}$, $c_{B,0}$, $\alpha$, $\beta$ within which spinodal decomposition occurs or our system remains stable; in other words, phase separation will not occur if the equilibrium volume fraction of component $C$ lies outside the spinodal region of the free energy $(f''(\varphi_e)<0)$.

\begin{figure*}[h]
	\begin{center} \
	\begin{minipage}[h]{1\linewidth} 
		{\includegraphics[scale=1]{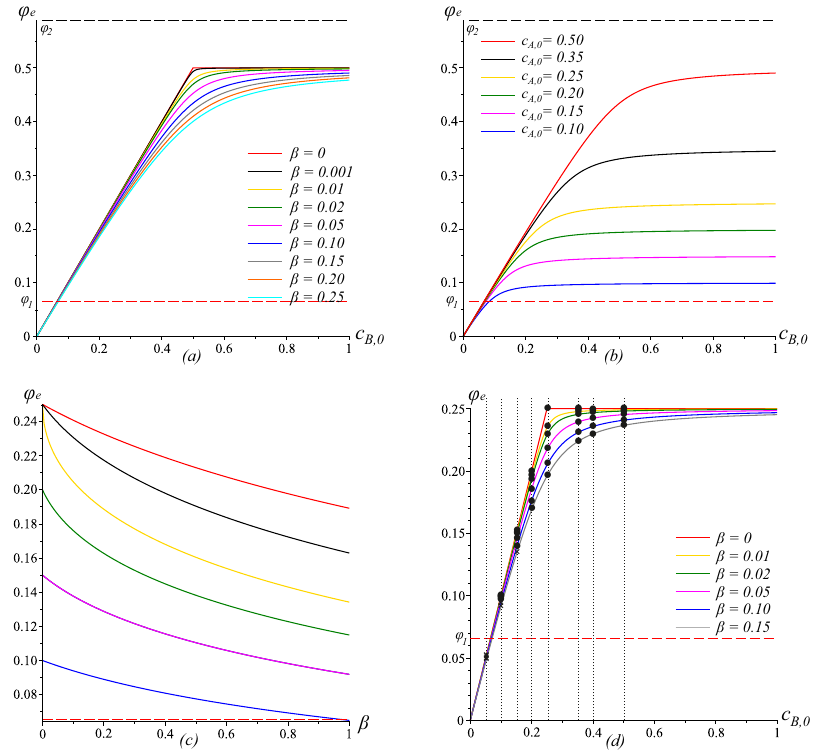}} 
	\end{minipage}
	\caption{Equilibrium concentration of component $C$ depending on the reaction parameters at $N=10$ and $\chi=1.3$ for the case: $(a)$ $c_{A,0}=0.25$, $(b)$ $\beta=0.1$, $(c)$ $c_{B,0}=0.25$ and $(d)$ $c_{A,0}=0.25$. The dashed lines correspond to $\phi_{1}$ (red) and $\phi_{2}$ (black). Plot (d) compares the analytical results with the numerical results: the black dots represent spinodal decomposition, while the black cross indicates the absence of it in the numerical simulations. \label{Fig}}\end{center}
\end{figure*}

Taking $N=10$ and $\chi=1.3$ in subsequent calculations, we obtain $\phi^s_{1,2}$ from \eqref{fi12}. 
The set of plots presented in Fig.~\ref{Fig} elucidates the relationship between the equilibrium volume fraction of component $C$ and the initial concentrations of organic molecules, phosphate ions and rate of the backward chemical reaction. 

Keeping $c_{A,0}$ constant, we studied the change in $\phi_e$ with respect to $c_{B,0}$ for different values of $\beta$, as shown in Fig.~\ref{Fig}(a).
A remarkable observation emerges: in a certain range of parameters on the graph, there is a region where the curves consistently lie below the concentration $\phi^s_{1}$ of the dense phase. This phenomenon indicates that spinodal degradation is excluded in this region, emphasizing the need for a minimum phosphate concentration to initiate spinodal degradation.
Furthermore, an intriguing trend was observed: curves associated with higher values of $\beta$ descended onto the plot, implying that an increased concentration of $c_{B,0}$ was required for spinodal decomposition. Additionally, an increase in $c_{A,0}$ (see Appendix) corresponds to a reduced requirement of $c_{B,0}$ for phase separation to occur. These findings provide valuable insight into the interplay between the initial concentrations and dissociation coefficient that govern phase separation in the system.

Analogous results were obtained for another set of curves shown in Fig.~\ref{Fig}(b). Here, we explore the relationship between $\phi_e$ and $c_{B,0}$ for different values of $c_{A,0}$, while maintaining a constant coefficient of the backward reaction, $\beta$. When fixing parameter $\beta$, our observations revealed that lower concentrations of $c_{A,0}$ required higher concentration of $c_{B,0}$ to induce spinodal decomposition.

The comparison between the analytical results obtained from equations \eqref{fi12}-\eqref{f_eq} and the numerical results obtained using COMSOL Multiphysics software revealed that the theoretical predictions exhibited a wider range of parameters for spinodal decomposition than the numerical calculations, as shown in Fig.~\ref{Fig}(d).

\section{Stability Analysis}

Another approach to examine linear stability is to analyze whether small perturbations $(\delta c_A, \delta c_B, \delta \phi)$ to the equilibrium state of the system will grow or decay. By substituting $c_a=c_{A,e}+\delta c_A$, $c_B=c_{B,e}+\delta c_B$, and $\phi=\phi_e+\delta \phi$ into equations \eqref{ca_eq}-\eqref{phi_d} (for simplicity, omitting the hat notation), we obtain the following equations for small fluctuations:
\begin{gather} \label{ca-delta}
    \cfrac{\partial \delta c_A}{\partial t}=D_A \nabla^2 \delta c_A - \alpha (c_{A,e}\delta c_B +\delta c_A c_{B,e}) +\beta \delta \phi\\ \label{cb-delta}
    \cfrac{\partial \delta c_B}{\partial t}=D_B \nabla^2 \delta c_B - \alpha (c_{A,e}\delta c_B +\delta c_A c_{B,e}) +\beta \delta \phi \\ \label{fi-delta}
    \cfrac{\partial \delta \phi}{\partial t}= \nabla^2(f''(\phi_e)\delta \phi -\nabla^2 \delta\phi) + \alpha (c_{A,e}\delta c_B + \delta c_A c_{B,e})-\beta \delta \phi
\end{gather}

\begin{figure*}[h]
	\begin{center} \
	\begin{minipage}[h]{1\linewidth} 
		{\includegraphics[scale=1]{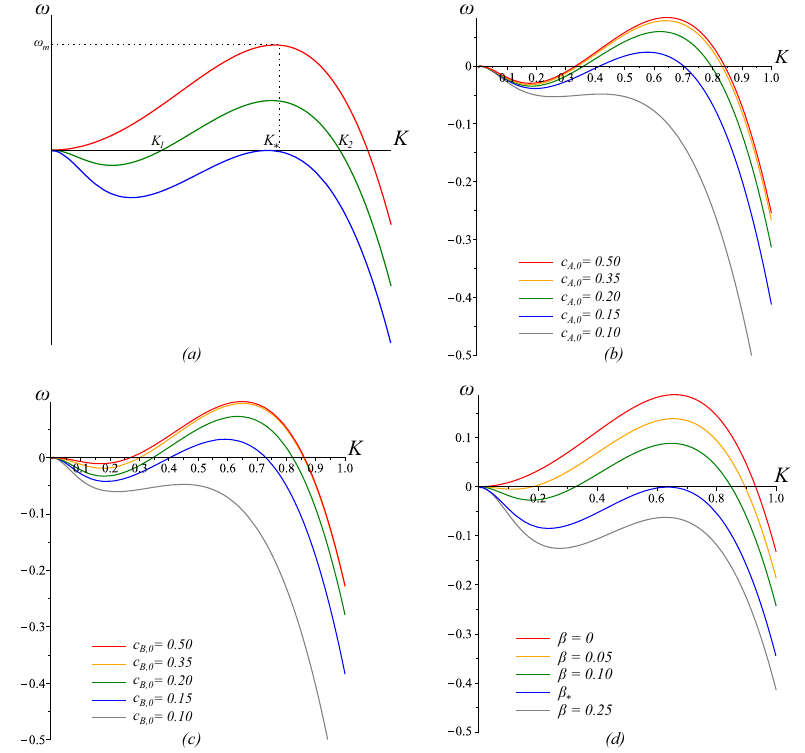}} 
	\end{minipage}
	\caption{Dispersion relation $\omega(K)$: $(a)$  three scenarios,  $(b)$ and $(c)$ for different initial concentrations of components $A$ and $B$, respectively,
at $\beta=0.1$ and for $(b)$ $c_{B,0}=0.25$, for $(c)$ $c_{A,0}=0.25$; $(d)$ for different values of $\beta$
 and $c_{A,0}=c_{B,0}=0.25$ revealing the strong impact of the dissociation constant on the spinodal decomposition. Shared parameters: $D_A=100$, $D_B=1000$, $\alpha=10$.\label{w(k)}}\end{center}
\end{figure*}

We seek a solution of the form:
\begin{gather}
    \delta c_A=h_A \exp{\left(i(k_1 x+ k_2 y) + \omega t\right)} \nonumber \\
    \delta c_B=h_B \exp{\left(i(k_1 x+ k_2 y) + \omega t\right)} \nonumber \\
    \delta \phi=h_C \exp{\left(i(k_1 x+ k_2 y) + \omega t\right)} \nonumber
\end{gather}

The aim is to obtain a separable solution. By substituting the ansatz into equations \eqref{ca-delta}-\eqref{fi-delta}, we obtain a set of equations for the amplitudes, which can be expressed in matrix form (2):
\begin{gather}
    \begin{pmatrix} \omega + K^2 D_A +\alpha c_{B,e} & \alpha c_{A,e} & -\beta \\ \alpha c_{B,e} & \omega +K^2 D_B +\alpha c_{A,e} & -\beta\\ -\alpha c_{B,e} & -\alpha c_{A,e} & \omega + K^2(f''(\phi_e) +K^2) +\beta \end{pmatrix} \begin{pmatrix}
    h_A\\h_B\\h_C
    \end{pmatrix} =0
\end{gather}
where $K^2=k_1^2+k_2^2$. To determine the solution to the linear system of homogeneous equations for the reaction-diffusion system, we calculate the determinant of the coefficient matrix. This results in a cubic equation for the growth rate $\omega$, which depends on the wave vector $K$ as follows:
\begin{gather} \label{w3}
    \omega^3 + \mathcal{G} (K^2)\omega^2 + \mathcal{P}(K^2) \omega + \mathcal{V}(K^2)=0
\end{gather}
where
\begin{gather}
    \mathcal{G}(K^2)=K^4 +\left(D_A+D_B+f''(\phi_e)\right)K^2 + \alpha(c_{A,e}+c_{B,e})+\beta\\
    \mathcal{P}(K^2)=K^2\left\{ (D_A+D_B)K^4 +\left((D_A+D_B)f''(\phi_e)  +D_A D_B +\alpha (c_{A,e}+c_{B,e})\right)K^2+ D_A(\alpha c_{A,e} +\beta)  \right. \nonumber \\ \left.  +D_B (\alpha c_{B,e} +\beta) +\alpha(c_{A,e}+c_{B,e})f''(\phi_e) \right\} \\
    \mathcal{V}(K^2)= K^4 \left\{ D_A D_B(K^4+K^2 f''(\phi_e)+\beta)  +\alpha(D_a c_{A,e}+D_Bc_{B,e} )(K^2+f''(\phi_e)) \right\}
\end{gather}

All three solutions of the equation are real in the region of interest within the parameter space. Two of these solutions consistently yielded negative values across all wave vectors.  The remaining solution, distinguished by positive values, is referred to the dispersion relation $\omega(K)$ that determines the growth rates for each mode with wavenumber $K$. 
In Fig.~\ref{w(k)}(a), distinct scenarios of the characteristic growth rate are presented, each represented by a unique curve. 

For wavenumber values $K$ where $w(K)>0$, the system becomes unstable, initiating spinodal decomposition.
In general, the function $w(K)$ transitions to negative values beyond a critical wavenumber $K_*$, marked by $\omega(K_*)=0$.  The blue curve in Fig.~\ref{w(k)}(a) represents a critical case in which the curve is entirely below the horizontal axis ($\omega=0$). Interestingly, it features a point of tangency with the $K$-axis, highlighting an equilibrium point at $\omega(K_*)=0$ in the dispersion relation. 
In addition, it should be noted that fluctuations characterized by wavelengths shorter than $\lambda_*=2\pi/K_*$ decay with time.

In this particular case, the model parameters must meet two essential conditions: first, they must satisfy equation \eqref{w3}, and second, the derivative of $\omega$ with respect to $K^2$ must equal zero, expressed as $\cfrac{d\omega}{dK^2}=0$.
This implies that $\mathcal{V}$ and $\mathcal{V}'_{K^2}$ must simultaneously be equal to zero.
We obtain the following quadratic equation for $K^2$:
\begin{gather}
 K^4 + \left( f''(\phi_e)+\alpha\left(\cfrac{c_{A,e}}{D_B} + \cfrac{c_{B,e}}{D_A}  \right)\right)K^2  + \alpha \left(\cfrac{c_{A,e}}{D_B}+ \cfrac{c_{B,e}}{D_A} \right) f''(\phi_e) +  \beta =0
\end{gather}
When this equation possesses a repeated root, it introduces the specific conditions necessary to attain patterning and determine the critical wavenumber as follows:
\begin{equation*}
  \left\{
    \begin{aligned}
      &  K^2_* = - \cfrac{1}{2} \left(\alpha \left( \cfrac{c_{A,e}}{D_B} + \cfrac{c_{B,e}}{D_A}\right) +f''(\phi_e) \right) \\
      &\left( \alpha\left(\cfrac{c_{A,e}}{D_B} + \cfrac{c_{B,e}}{D_A} \right) -f''(\phi_e)\right)^2=4\beta \\
      &  -f''(\phi_e) > \alpha \left( \cfrac{c_{A,e}}{D_B} + \cfrac{c_{B,e}}{D_A} \right) 
    \end{aligned}
  \right.
\end{equation*}

In the second curve in Fig.~\ref{w(k)}(a), depicted in green, we observe a scenario in which a portion of the plot extends above zero. This behaviour results in two distinct roots corresponding to the exact points at which the curve intersects the $K$-axis. 
Moreover, when the condition $\left(\alpha\left(\cfrac{c_{A,e}}{D_B} + \cfrac{c_{B,e}}{D_A}\right) - f''(\phi_e)\right)^2 > 4\beta$ is satisfied, this signifies the existence of two roots for $\mathcal{V}(K^2)=0$. In simpler terms, this condition implies the presence of two specific values, denoted as $K_{1,2}$, for which $\omega(K_{1,2})=0$:
\begin{gather}
    K^2_{1,2}= - \cfrac{1}{2} \left\{ f''(\phi_e) + \alpha \left( \cfrac{c_{A,e}}{D_B} + \cfrac{c_{B,e}}{D_A} \right)  \pm \sqrt{\left( 
\alpha \left( \cfrac{c_{A,e}}{D_B} + \cfrac{c_{B,e}}{D_A}\right) - f''(\phi_e) \right)^2 -4 \beta} \right\}
\end{gather}

The characteristic size of the initially emerging phase regions is defined by the wavelength $\lambda_{max}=2\pi/K_{max}$, where $K_{max}$ is the wave vector at which $\omega(K)$ reaches its maximum (red curve in Fig.~\ref{w(k)}(a)), which determines the wavelength of the fastest growing Fourier mode.  
To obtain the case for the maximum growth rate, we must write the growth equation derived from \eqref{w3}. After simplifications, it takes the following form:
\begin{gather}
\omega(K^2)=\mathcal{H}(K^2)- \mathcal{J}\left(\alpha(c_{A,e}+c_{B,e}) +\beta\right)
\end{gather}
where $\mathcal{H}(K^2)$ represents a function of $K^2$, and $\mathcal{J}$ denotes a constant term. Therefore, in the absence of chemistry ($\alpha=\beta=0$), the growth factor aligns with the conventional prediction of Cahn's linear theory. The simultaneous presence of a reaction leads to a decrease in the typical growth rate by an amount proportional to the combined influence of the forward and reverse reaction rates, denoted $\alpha$ and $\beta$. This setting shifts the threshold for small wavelengths toward larger values and introduces a threshold for long wavelengths. Consequently, the concentration fluctuations at large wavelengths (small $K$) were restricted by the influence of the reactions. The suppression of long wavelengths is a natural mechanism for pattern selection in various systems.

If $\left(\alpha\left(\cfrac{c_{A,e}}{D_B} + \cfrac{c_{B,e}}{D_A} \right) - f''(\phi_e)\right)^2 < 4\beta$ then patterning cannot occur. On the other hand, if the condition for pattern formation:
\begin{gather} \label{pattern_cond}
    \left(\alpha\left(\cfrac{c_{A,e}}{D_B} + \cfrac{c_{B,e}}{D_A} \right) - f''(\phi_e)\right)^2 \geq 4\beta
\end{gather}
is satisfied, then patterning can occur.

In our stability analysis, we examine the impact of the initial concentrations $c_{A,0}$ and $c_{B,0}$ of components $A$ and $B$ (with $\phi=0$) and assume a rapid establishment of the equilibrium of the reaction according to \eqref{f_eq}.
Fig.~\ref{w(k)} provides illustrative examples of the dispersion relation for various reaction parameters. 

In Fig.~\ref{w(k)}$(b),(c)$, we explore the influence of the initial concentrations $c_{A,0}$ (organic molecules) and $c_{B,0}$ (phosphate ions), respectively, on the dispersion relation while keeping other parameters constant. The curves highlight that a sufficiently high concentration of organic molecules and phosphate ions is necessary to enable spinodal decomposition by entering a wave vector region where $\omega(K)>0$. Moreover, the wavelength of the fastest-growing Fourier mode showed a slight decrease with increasing initial concentration.

The change in the dispersion relationship with increasing dissociation constant $\beta$ is shown in Fig.~\ref{w(k)}$(d)$. When dissociation is negligible $(\beta=0)$, the frequency $\omega(K)$ remains positive for all wave vectors $K<K_*$. This suggests that even long-wavelength fluctuations experience growth, leading to the continuous coarsening of the phase regions during the linear growth regime. Consequently, the formation of regular stationary phase patterns is hindered. However, an interesting phenomenon arises with increasing dissociation constant $\beta$. Within a specific range of small wave vectors, the growth rate becomes negative, resulting in the suppression of the long-wavelength fluctuations. This suppressed wave vector region is crucial for establishing regular stationary phase patterns and preventing the continuous coarsening of phase regions. As the dissociation constant increases further, the growth rate $\omega(K)$ becomes negative for all wave vectors, indicating that spinodal decomposition cannot occur.

It is crucial to note that the stability analysis presented herein predominantly focuses on the initial stages of phase separation. Other methods, such as numerical computations, are required to comprehend the subsequent nonlinear evolution of concentration. Nevertheless, this analysis offers significant insights into the identification of parameter combinations that are conducive to the formation of regular stationary phase patterns.

\section{Numerical analysis of phase separation}

\subsection{1D simulations}

To provide insight into the nonlinear phase separation process, one-dimensional numerical simulations were conducted. As an initial condition, the concentrations $c_{A,0}$ and $c_{B,0}$ of components $A$ and $B$ were assigned with small fluctuations of the order of 1\%. These initial fluctuations decay quickly as a result of diffusion processes. The initial concentration of component $C$ was chosen to be very low ($\phi = 0.002$). We applied periodic boundary conditions assuming a significantly higher diffusion coefficient for phosphate ions than for organic molecules. This ensured that the diffusion processes were faster than the phase separation process. Specifically, the diffusion coefficients of $D_A=100$ and $D_B=1000$ for components $A$ and $B$, respectively, were chosen. Furthermore, the reaction constant $\alpha$ was maintained at a constant value of $\alpha=10$.

Before proceeding with further analysis, it is important to note that, in 1D, reaction-diffusion systems can exhibit two types of patterns and their combinations: peak and mesa patterns.

\begin{figure*}[h]
	\begin{center} \
	\begin{minipage}[h]{1\linewidth} 
		{\includegraphics[scale=1]{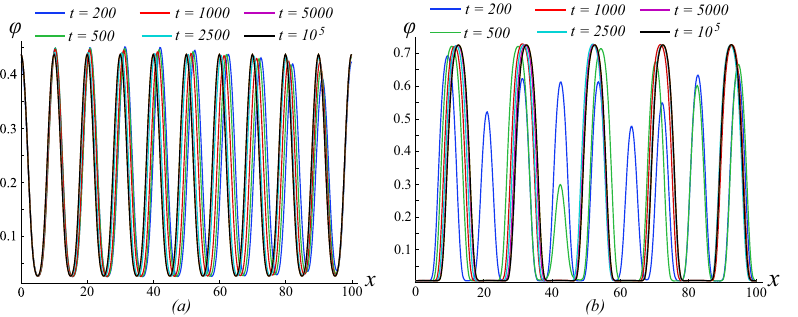}} 
	\end{minipage}
	\caption{1D concentration profile of component $\phi$ at various time. Parameters: $c_{A,0}=c_{B,0}=0.25$, (a): $\beta=0.1$, (b): $\beta=0.01$.\label{1d_time}}\end{center}
\end{figure*}

Fig.~\ref{1d_time} illustrates the evolution of the concentration of component $C$ over time. In Fig.~\ref{1d_time}(a), the system exhibits dynamic characteristics of peak-forming patterns when $\beta=0.1$. This behaviour is characterized by the formation of a sequence of density peaks, followed by a slow coarsening or shrinking process. Notably, the dynamics of each peak highlight the competition for mass among neighboring peaks, as the mass redistributes between pattern domains. However, over time it becomes evident that the observed patterns reach a stationary configuration. Notably, the number of peaks remains constant throughout the process.

The coarsening process, predominantly driven by both competition and coalescence, is shown in Fig.~\ref{1d_time}(b) for $\beta=0.01$. The coalescence dynamics is governed by mass competition within the low-density regions, where mass competition occurs between the "troughs" of the patterns. As a peak moves toward a neighboring peak, one trough grows, whereas the other collapses. Consequently, both the competition and coalescence processes are propelled by the destabilizing distribution of mass between domains of high or low density. In other words, mass-competition instability arises between the domains with varying densities.

By examining these two plots, it is apparent that for smaller values of the dissociation constant the coarsening process takes place, whereas an increase in the dissociation constant leads to shorter wavelengths in the patterns.

\begin{figure}[h]
	\begin{center} \
	\begin{minipage}[h]{1\linewidth} 
		{\includegraphics[scale=1]{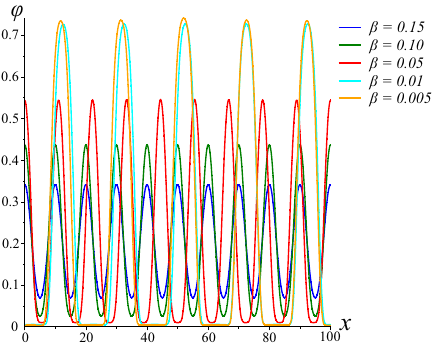}} 
	\end{minipage} 
		\caption{1D concentration profile of component $C$ at various $\beta$ and $t=10^5$,  $c_{A,0}=c_{B,0}=0.25$ \label{c(x)_b}}
	\end{center}
\end{figure}

The effect of the dissociation constant on the evolution of the concentration is shown in Fig.~\ref{c(x)_b}. The concentration profiles depict the profound influence of the varying dissociation constant on the dynamics of the phase separation, underscoring its critical role in this process. A regular arrangement of concentration peaks is quickly established. Notably, an increase in the value of $\beta$ leads to higher frequencies and smaller amplitudes of the concentration waves.

\begin{figure*}[h]
	\begin{center} \
	\begin{minipage}[h]{1\linewidth} 
		{\includegraphics[scale=1]{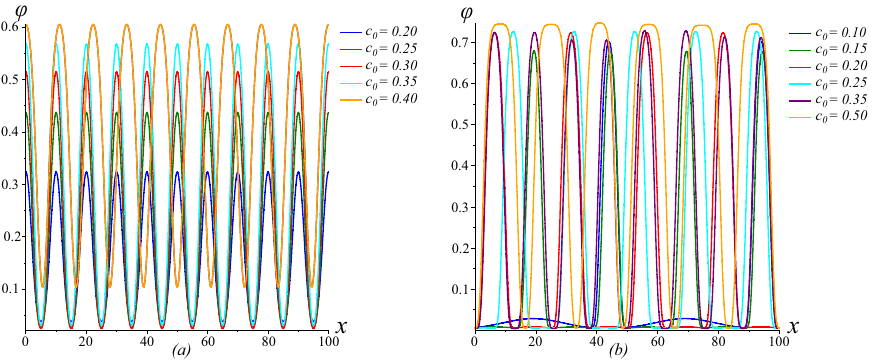}} 
	\end{minipage}
	\caption{1D concentration profile $\phi$ at various initial concentrations $c_{A,0}=c_{B,0}=c_0$ at $t=10^5$, (a): $\beta=0.1$, (b): $\beta=0.01$.\label{c(x)_c}}
	\end{center}
\end{figure*}

Similarly, Fig.~\ref{c(x)_c} illustrates the one-dimensional volume fraction profile of component $C$ for different initial concentrations at two different dissociation constant values. Fig.\ref{c(x)_c}(a) shows the dynamics of the peak-forming patterns for $\beta= 0.1$. Evidently, increasing the initial concentrations of the reacting components results in longer wavelengths and greater amplitudes. In contrast, 
as demonstrated in Fig.~\ref{c(x)_c}(b), as the dissociation constant $\beta$ decreases, the system undergoes a change in dynamics. For low initial concentrations, only a few peaks with extensive trough areas are observed. However, as the concentration increases, the trough regions decrease in size, leading to the formation of mesa patterns characterized by increased crest areas.

A comparative analysis of the results allow us to draw conclusions regarding the influence of the dissociation constant on the concentration patterns. Specifically, for a larger dissociation constant ($\beta=0.1$), the concentration peaks exhibit a regular and rapid establishment. In contrast, for lower values of $\beta$ ($\beta=0.01$), coarsening of the phase region is observed.

The concentration profiles obtained in Fig.~\ref{c(x)_b} and Fig.~\ref{c(x)_c} demonstrate stationary behaviour as they remain unchanged throughout the calculation period from $t=10^4$ to $t=5\cdot10^5$. The wavelengths of the regular concentration profiles approximately correspond to the wave vector at the maximum $K_{max}$ of the dispersion relation, as shown in Fig.~\ref{w(k)}. These one-dimensional calculations provide valuable insights into how the choice of model parameters can influence the phase separation process in higher dimensions, highlighting important trends.

\subsection{2D simulations}

This section examines 2D simulations of concentration profiles. Previous studies have primarily focused on the use of 2D models to investigate the development of diatom structures and pattern formation. This choice was motivated by the computational ease and flat shape of the silica deposition vesicles in diatoms. In this section, we present the findings of 2D simulations that demonstrate steady state patterns in the volume fraction of component $C$. Initially, we simulate the system of equations \eqref{ca_eq}-\eqref{phi_d} in a square domain with periodic boundary conditions. 
As in the 1D results, we fixed the domain size to be 100, the diffusion coefficients to be $D_A=100$ and $D_B=1000$, and the phosphorylation rate $\alpha$ to be 10. The calculation period for the simulations is $t=5\cdot 10^5$.

\begin{figure*}[h]
	\begin{center} \
	\begin{minipage}[h]{1\linewidth} 
		{\includegraphics[scale=1]{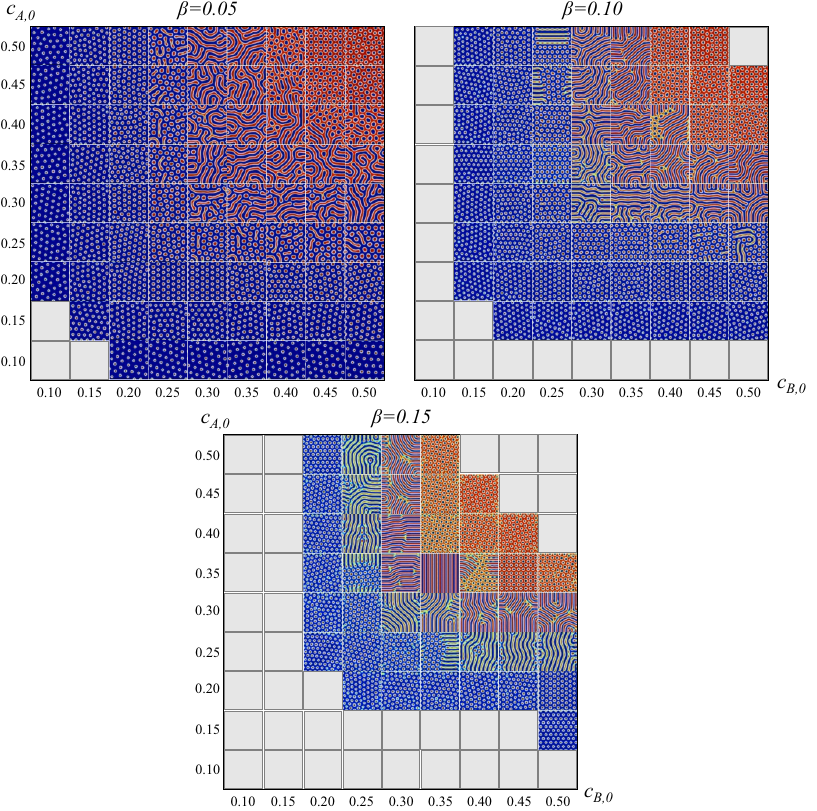}} 
	\end{minipage}
 \caption{Concentration profile at fixed dissociation rate \label{fig_fixed_b}, the red color corresponds to the organic-rich phase. For the maximum concentration see Fig.~\ref{color_bar}.}
	\end{center}
\end{figure*}

\begin{figure*}[h]
	\begin{center} \
	\begin{minipage}[h]{1\linewidth} 
		{\includegraphics[scale=0.75]{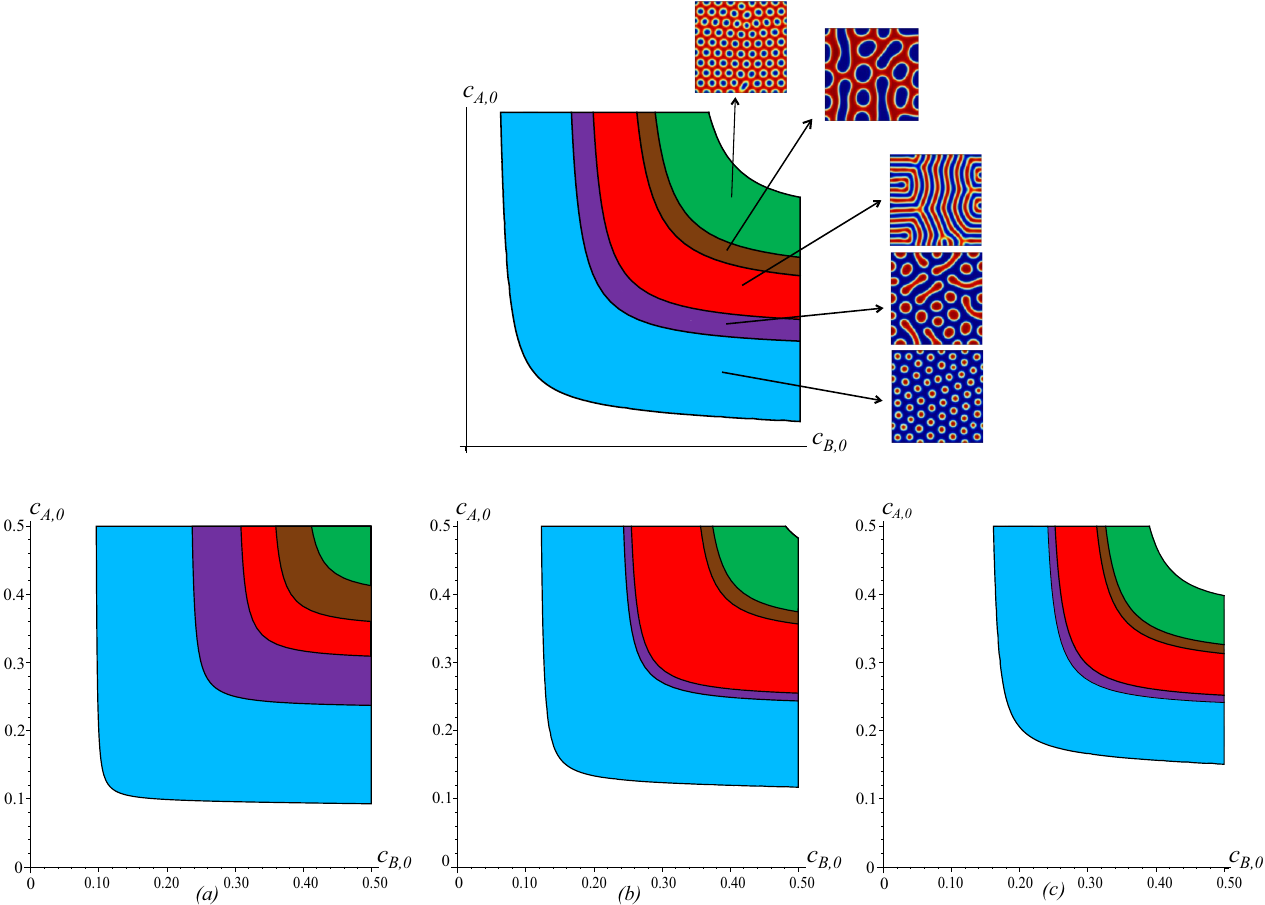}} 
	\end{minipage}
	\caption{Pattern diagram at fixed values of dissociation rate and and particular cases for given values: $(a)$ $\beta=0.05$, $(b)$ $\beta=0.10$ and $(c)$ $\beta=0.15$.\label{beta_fixed_scheme}}
	\end{center}
\end{figure*}

In the first set of simulations, we maintained a constant dissociation rate $\beta$ (set to a specific value e.g for each illustrated data set in Fig.~\ref{fig_fixed_b}) for simplicity while varying the initial concentrations of the components $A$ and $B$. This was performed to observe 
how these concentrations affect the formation of the patterns. The results for different values of the dissociation rate $\beta$ are shown in Fig.~\ref{fig_fixed_b}, where the organic-rich phase is represented in red. The simulations demonstrate that the system exhibits five different types of patterns, as illustrated in the schematic pattern diagram and particular cases for the given values of $\beta$ in Fig.~\ref{beta_fixed_scheme}. Among these types, three are considered as the main types, while the remaining two are 'transitional' or 'mixed' types.

The maximum concentration varies with the parameter $\beta$,  see Fig.~\ref{color_bar}.

The first type of patterns, shown in blue in Fig.~\ref{fig_fixed_b}, occurs when the initial concentrations of phosphate ions and organic molecules are low.  These patterns exhibit a regular structure and consist of aggregates of organic "droplets" that are uniformly sized. The arrangement of these droplets is hexagonal or nearly hexagonal.

The second category of patterns, represented by the violet color in the pattern diagram, emerges when the initial concentrations are increased. In this case, the organic phase expands, leading to the elongation of certain droplets and the formation of alignments. These patterns exhibit a combination of aligned structures and droplets.

As the concentration further increases, distinct patterns appear in lamellar (labyrinth-like) structures. These patterns indicate that no dominant phase is present. In Fig.~\ref{fig_fixed_b}, these patterns are shown in red. It should be noted that these stripes remain stationary despite their irregular structure.

When the initial concentrations are high, a regular structure of a dense network of organic-rich phase components $C$ with circular pores of organic-poor phase forms. These patterns are shown in green in the figure.

Finally, patterns characterized by a dense network consisting of a combination of stripes and pores of the organic-dilute phase are shown in brown in Fig.~\ref{fig_fixed_b}.

The influence of periodic boundary conditions on the observed patterns in our simulations is an important factor to consider because of the limited size of the simulation cell. To minimize the artificial effects caused by this limitation, we conducted simulations with periodic boundary conditions. This approach was chosen to optimize the simulations for constrained geometries using a phase separation mechanism. 

In diatoms, aggregation of organic material occurs within the Silica Deposition Vesicle (SDV). Notably, the confinement of the SDV is significantly larger than that of the pores observed in the diatom valves. Hence, it is reasonable to focus on a small section of the SDV and assume periodic boundary conditions in the calculations. This decision is in line with the practical limitations of the diatom structures.

Moreover, experimental evidence indicates that the type of soluble silica, together with the biomolecules involved in silica precipitation, can significantly influence the resulting silica morphology under various conditions \cite{lechner2015silaffins,sumper2006learning,bernecker2010tailored}.

\begin{figure*}[h]
	\begin{center} \
	\begin{minipage}[h]{1\linewidth} 
		{\includegraphics[scale=0.75]{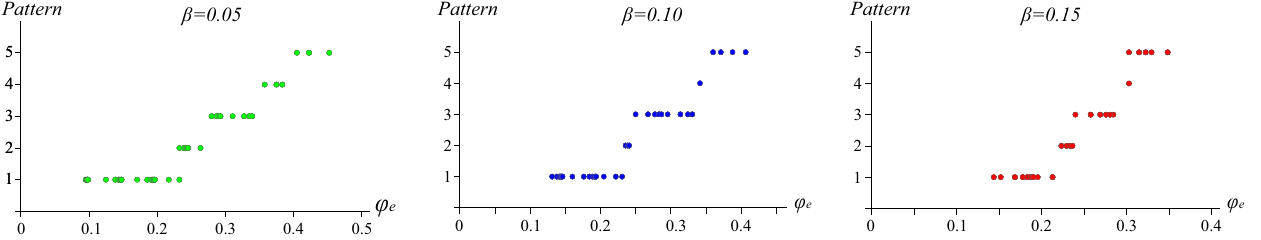}} 
	\end{minipage}
	\caption{Pattern's type vs equilibrium free energy $\phi_e$ at precise values of $\beta$.} \label{f_phi_beta}
	\end{center}
\end{figure*}

The region in which spinodal decomposition occurs is delineated by lines determined using equation \eqref{pattern_cond}. By computing the equilibrium free energy $\phi$ using equation \eqref{f_eq} for various combinations of parameters and plotting it against the corresponding pattern types, we observe that each pattern type emerges within a specific range of $\phi_e$ values. This correlation is illustrated in Fig.~\ref{f_phi_beta}.

\begin{figure*}[h]
	\begin{center} \
	\begin{minipage}[h]{1\linewidth} 
		{\includegraphics[scale=1]{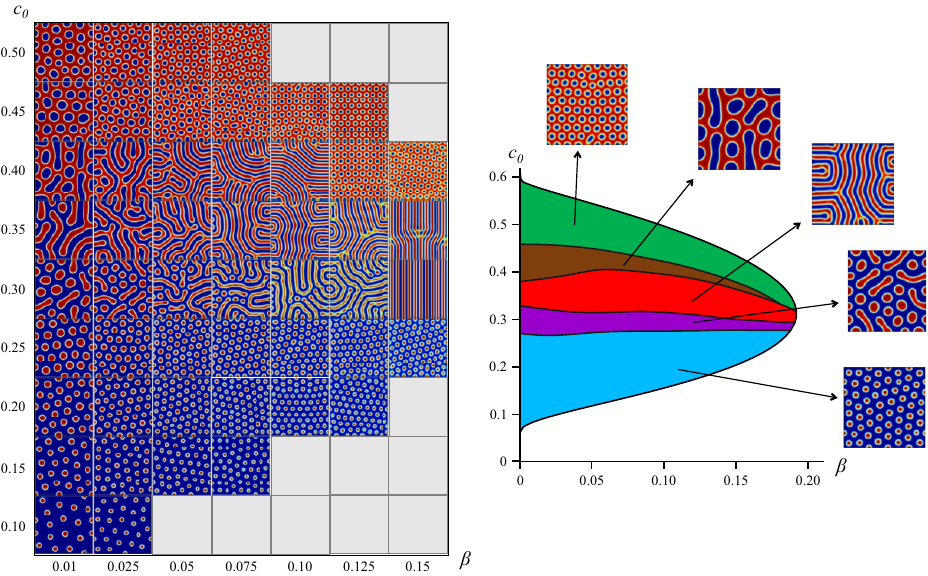}} 
	\end{minipage}
	\caption{Stationary patterns of 2D concentration of component $C$ up to very large computation time ($c_{A,0}=c_{B,0}=c_0$), the red color corresponds to the organic-rich phase (displayed on the left, for the maximum concentration see Fig.~\ref{color_bar}) and Pattern diagram: the grey color indicates a scenario where no pattern formation (shown on the right).}\label{square}\end{center}
\end{figure*}

In the second set of simulations, we set the initial concentrations of phosphate ions and organic molecules to be equal, denoting $c_{A,0}=c_{B,0}=c_0$. By varying $c_0$ and $\beta$, we obtained two-dimensional profiles of the concentration of component $C$, as demonstrated in Fig.~\ref{square}. This study provides valuable insights into the impact of the backward reaction and initial concentrations on pattern formation. The resulting pattern diagram is shown in Fig.~\ref{square}. Similarly to the previous set of simulations, we observe that the system exhibits five types of patterns.

Analyzing the results of the two sets of simulations provides valuable insights into the influence of initial concentrations and reverse reaction on pattern formation. 
An in-depth analysis of the patterns obtained in Fig.\ref{fig_fixed_b}, for the case where the dissociation rate remains constant, reveals that the parameter $\beta$, which represents the rate of the backward reaction, plays a key role in determining the wavelength of the resulting two-dimensional concentration profiles of component $C$. This suggests that for a given value of $\beta$, the minimum distance between the two nearest droplets in the droplet patterns, between the midlines for lamellar patterns or neighboring pores in the dense network of the organic-rich phase remains approximately constant regardless of changes in initial concentrations $c_{A,0}$ and $c_{B,0}$.

\begin{figure*}[h]
	\begin{center} \
	\begin{minipage}[h]{1\linewidth} 
		{\includegraphics[scale=0.9]{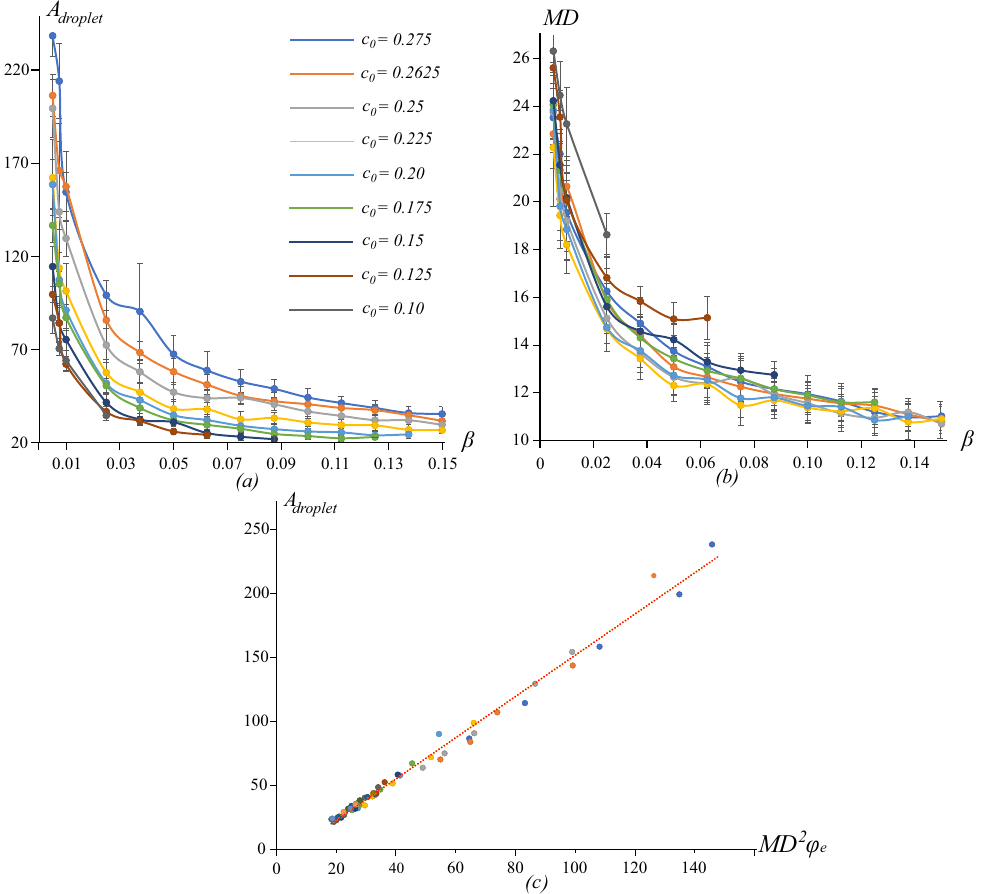}} 
	\end{minipage}
	\caption{(a) Average size of organic droplets depending on dissociation rate for various initial concentration. (b) Minimum distance between two closest neighboring droplets. (c) Pattern scaling: dependence of area of droplet on pattern wavelength and equilibrium volume fraction. \label{AS_MD_conv}}
	\end{center}
\end{figure*}

Moreover, the variations in $\beta$ shown in Fig.~\ref{square} cause either refinement or coarsening of the structure, depending on whether the dissociation rate increases or decreases, respectively. This correlation is illustrated in Fig.~\ref{AS_MD_conv}(a) for droplet patterns. An increase in $\beta$ results in a shorter wavelength pattern. This relationship is illustrated in Fig.~\ref{AS_MD_conv}(b), which graphically represents the average wavelength and its standard deviation. For fixed initial concentration the size of the droplets, pores and lamellar structures scale with the pattern wavelength so that decreasing $\beta$ essentially causes the entire pattern to expand, as shown in Fig.~\ref{AS_MD_conv}(c).

It is important to note that the similarities between the observed variations in $\beta$ and their impact on the wavelength are consistent with the findings derived from the dispersion relation \eqref{w3} and the outcomes of 1D simulations conducted for the examined equation system. This relationship is depicted in Fig.~\ref{w(k)}$(c)$ and the 1D simulations are shown in Fig.~\ref{c(x)_b}. The agreement in findings across these different analytical approaches emphasizes the reliability and coherence of our conclusions regarding the complex dynamics of the system under the influence of different dissociation rates.

After a detailed examination of the results obtained from both sets of simulations, a significant finding becomes apparent: the morphology of the patterns is related to the initial concentrations of both reactants, $c_{A,0}$ and $c_{B,0}$. The transition from one type of pattern to another occurs only with a sufficient change in the concentrations of both reacting components. The size of the phase-separated components is largely influenced by the initial concentrations of phosphate ions and organic molecules. When the dissociation rate remains constant, an increase in the concentration of either of the reacting components leads to the enlargement of the area occupied by the organic-rich phase. This is illustrated in Fig.~\ref{AS_MD_conv}$(a)$. The expansion of the organic-rich area, in turn, helps in the creation of various organic-rich networks that exhibit unique characteristics specific to each pattern. It is important to mention that the rate at which the organic-rich phase grows is higher when the initial concentration of phosphate ions increases, compared to the same increase in the concentration of organic molecules.
Based on these observations, we can identify a method for adjusting the scale of the patterns, as shown in Fig.~\ref{AS_MD_conv}$(c)$ for the droplet pattern.

\begin{figure}[h]
	\begin{center} \
	\begin{minipage}[h]{1\linewidth} 
		{\includegraphics[scale=1]{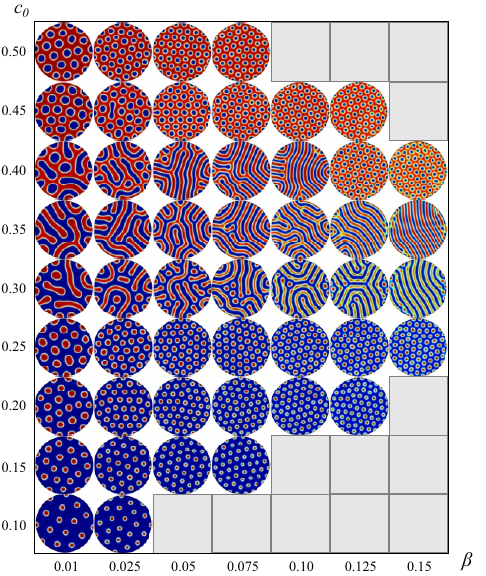}} 
	\end{minipage}  
		\caption{Circular stationary patterns: Concentration profiles of component $C$ within a circular boundary ($c_{A,0}=c_{B,0}=c_0$). For the maximum concentration, refer to Fig.~\ref{color_bar}. \label{circle}}
	\end{center}
\end{figure}

To study the formation of patterns in the growing valve \cite{heintze2022molecular,de2023exocytosis} and the hierarchical pore structure of diatoms \cite{willis2010discrete,willis2013simple}, where larger pores act as confinements for the formation of smaller pores, a series of simulations were conducted using a circular geometry. To evaluate the stability and applicability of our results, simulations were performed in a circular domain with zero-flux boundary conditions. The resulting patterns are shown in Fig.~\ref{circle}. The obtained patterns remain stationary throughout the simulation from $t=2\cdot10^4$ to $t=5\cdot10^5$. However, it is important to note the possibility of changes occurring over extremely long periods.

It is worth mentioning that the patterns obtained within the circular simulation cell closely resemble those derived from previous sets of simulations, as shown in Fig.~\ref{fig_fixed_b},\ref{square}, when the model parameters have the same values. The uniformity across various geometries not only enhances the reliability of our findings but also emphasizes the inherent characteristics of the observed patterns, which are influenced by the dynamics governed by the specified model parameters.

\begin{figure*}[h]
	\begin{center} \
	\begin{minipage}[h]{1\linewidth} 
		{\includegraphics[scale=0.75]{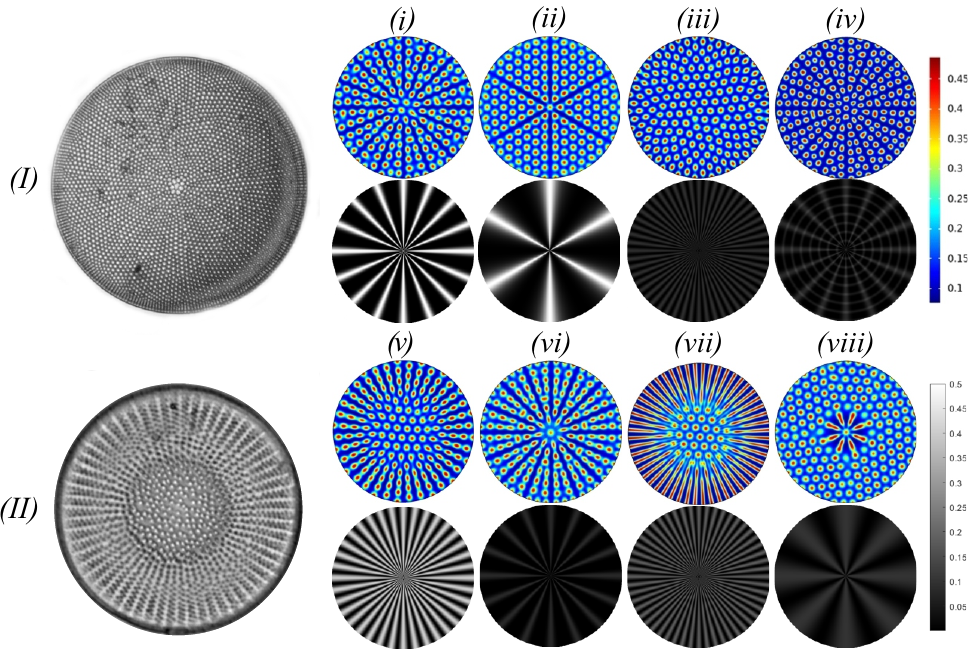}} 
	\end{minipage}
		\caption{Experimental (on the left) and simulated (on the right) structures of centric diatoms. (I) \textit{Coscinodiscus}, (II) \textit{Stephanodiscus} \textit{medius}.  During the simulations, the shared parameters were $\beta=0.15$, $c_{A,0}=0.2$, and $c_{B,0}=0.25$. Simulated images (top) and the corresponding prepatterning fields $h$ that were used in the simulations: (i) $0.5cos^{10}(8\theta)$, (ii) $0.5(1+10sin^2(3\theta))^{-1}$, (iii) $0.09cos^2(26\theta)$, (iv) $0.05cos^6(0.12 \pi \sqrt{x^2+y^2})+0.08cos^8(8\theta)$, (v) $0.4cos^2(18\theta)$, (vi) $0.08cos^8(8\theta)$, (vii) $0.18cos^2(26\theta)$, (viii) $0.1cos^2(4\theta)$. Permission for experimental images has been requested \cite{Diatom1,Diatom2}.} \label{prepattern_h_var}
	\end{center}
\end{figure*}

\begin{figure*}[h]
	\begin{center} \
	\begin{minipage}[h]{1\linewidth} 
		{\includegraphics[scale=0.5]{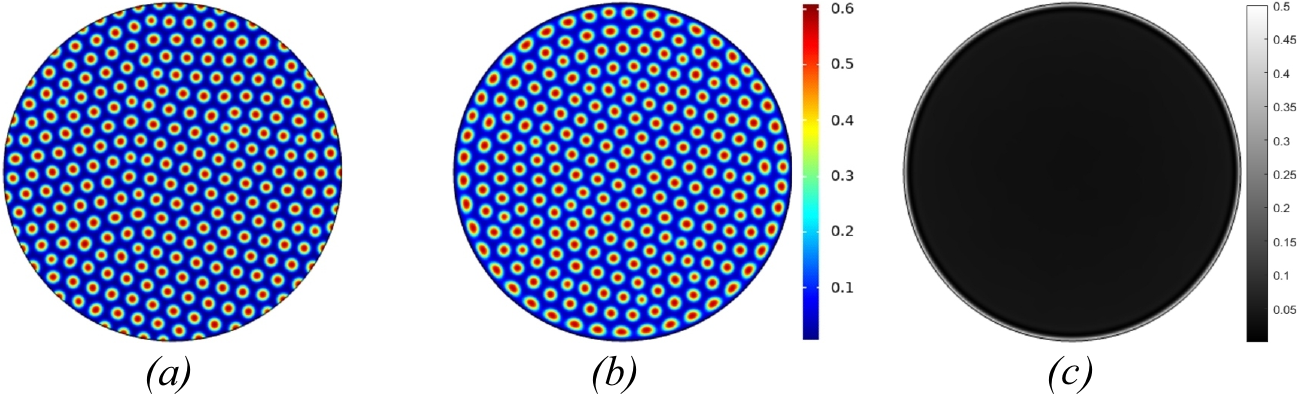}} 
	\end{minipage} 
		\caption{Effect of Prepatterning Field: (a) Pattern on a computational domain with zero-flux boundary conditions, (b) Pattern on a computational domain with corresponding prepatterning field on (c), which is defined by $h=0.5$ for $R\in[99.75,100]$ or $h=0$ otherwise. Shared parameters: $\beta=0.1125$, $c_{A,0}=c_{B,0}=0.225$.} \label{pore}
	\end{center}
\end{figure*}

\begin{figure*}[h]
	\begin{center} \
	\begin{minipage}[h]{1\linewidth} 
		{\includegraphics[scale=1]{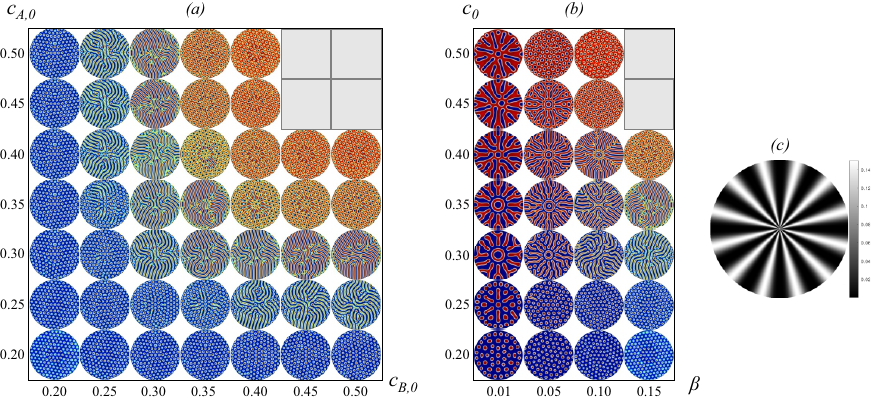}} 
	\end{minipage}
	\caption{Circular stationary patterns: Concentration profiles of component $C$ within a circular boundary for $(a)$$\beta=0.15$ and $(b)$ $c_{A,0}=c_{B,0}=c_0$ for  prepatterning field $h=0.15 sin^4(6\theta)$ on $(c)$. For the maximum concentration see Fig.~\ref{color_bar}.}\end{center} \label{prepattern_var_c}
\end{figure*}

This comprehensive study significantly enhances our understanding of the complex dynamics that govern the formation of patterns in the studied system. 
The observed patterns remain stationary during the simulation  for times after $t=2\cdot10^4$ 
and act as a template to guide silica deposition.

\section{Prepatterning}

In this section, our goal is to examine the impact of "prepatterning" caused by the existence of silica costae in the base layer on the liquid-liquid phase separation process. We extend the research conducted by Lenoci et al.\cite{lenoci2006self}  to investigate how prepatterns, such as costae that pre-exist the pores, impact the final patterns. Notably, in the work of Lenoci and Camp the final patterns were not stationary and were not coupled with chemical reactions.

To simulate the presence of silica costae, a characteristic feature of diatoms, during the early stages of frustule formation, we introduce an additional term $h$ into the equation of free energy \eqref{free_energy}. This term represents the local field generated by the pre-existing silica ribs, which contributes to the formation of the prepatterning structure. 
Thus, the modified free energy equation \eqref{free_energy} that governs the thermodynamic state of the phase-separating component becomes the following:
\begin{gather}
    F[\phi]=\int\left( f(\phi)+\cfrac{\kappa}{2} (\nabla\phi)^2+h\phi\right)
\end{gather}

The prepatterning field is designed to mimic the morphology of the costae observed in real diatoms. We focus on centric diatoms with circular valves, which can be classified into two types based on valve morphology. The first type includes diatoms whose valve faces exhibit a single common morphology. In this type, the areolae, which are regularly repeated pores, may be hexagonal, radiate from a central annulus, a hyaline ring, or be arranged in radial rows. 
Examples include Actinocyclus and Eupodiscus. The second type id comprised of diatoms with valves divided into two parts (central and marginal areas), each with distinct morphologies (e.g. central areolae surrounded by radiating striae). Examples of these include Brevisira and Stephanodiscus.

Based on this division and observation, we select appropriate functions for the prepatterning field to correspond to the shape of the costae. In Fig.~\ref{prepattern_h_var}, we present experimental images of selected diatoms from two distinct classes alongside a variety of simulated structures produced by employing different prepatterning fields, illustrating the variety of attainable patterns and highlighting how the prepatterning field can accurately represent the unique characteristics of the diatom morphology.

In addition, prepatterning fields can be used to solve the numerical problems encountered during the simulations. During our computations, we observed the formation of organic half-droplets at the boundaries of the numerical domain. To circumvent this problem and ensure that all droplets remain inside the simulation cell, we applied a thin layer of the prepattern field to the boundary. The numerical domain with zero-flux boundary
conditions is illustrated in Fig.~\ref{pore}(a), where half-droplets are visible at the boundaries.
In Fig.~\ref{pore} (b,c) for the same domain but with the inclusion of the prepatterning field at the boundary, all droplets are held within the cell. Importantly, all computations were performed using the same parameters, underscoring the effectiveness of the prepatterning field in eliminating numerical anomalies while maintaining the validity of the simulation outcomes.

Simulations with different initial concentrations and dissociation rates were conducted similarly to the previous sections. The results shown in Fig.~\ref{prepattern_var_c} elucidate how variations in the initial concentrations and dissociation rate affect the resulting patterns in the presence of prepatterning induced by silica costae.

Overall, incorporating the prepatterning field into our simulations not only enables us to precisely mimic the morphology of diatoms but also enhances the precision and robustness of our computational models by mitigating boundary effects and ensuring consistency with experimental observations.

\section{Discussion}

Our investigation into the role of phosphate ions and organic molecules in the formation of the diatom silica structure provided valuable insights into the mechanisms driving pattern formation. By exploring the hypothesis of organic template-driven silica morphogenesis, our study focused on the aggregation of organic molecules through a phase separation model, supported by \textit{in vitro} experiments highlighting the catalytic role of organic components.

The approach used in this study couples the phase separation process with a chemical reaction, allowing the phase-separating component to dissociate into two constituents. Unlike previous studies \cite{bobeth2020continuum,lenoci2006self}, our work involved detailed analysis to elucidate the conditions necessary for pattern formation and to understand how model parameters and prepatterning field affect pattern morphology and transitions between different types of patterns. 

The study in \cite{glotzer1995reaction} shows that the chemical reaction suppresses spinodal decomposition, limiting the ongoing coarsening of the phase domains during phase separation. Building on the framework presented by Bobeth et al.\cite{bobeth2020continuum}, which couples phase separation with chemical reactions, the objective of this study was to explore the effects of model parameters on pattern formation. We provided valuable insights into the interplay between the initial concentrations and dissociation coefficients that govern phase separation. Through stability analysis, we established conditions for pattern formation and assessed how the initial concentrations and dissociation rate affect the dispersion relation. Our numerical simulations identified five distinct pattern types and emphasised the role of the initial concentrations and the reverse reaction rate in the formation of these patterns. Experimental evidence indicated that the type of soluble silica and biomolecules significantly influence silica morphology under different conditions \cite{bernecker2010tailored,lechner2015silaffins,sumper2006learning}. Consequently, our findings provide a robust framework for integrating these insights into more advanced models and \textit{in vitro} experiments designed to clarify these morphological variations in silica. In our work, we demonstrated that the pattern morphology depends on the initial concentrations of organic molecules and phosphate ions. We observed that transitioning between pattern types requires significant changes in the concentrations of both reacting constituents. Increasing the concentration of any reactant enlarges the region dominated by the organic-rich phase, facilitating the formation of diverse organic-rich networks, each exhibiting distinct characteristics that reflect their pattern types.

In \cite{lenoci2006self}, phase separation was induced by an additional local field due to the existing silica costae in the base layer, leading to the formation of various two-dimensional patterns that closely resembled the observed valve structures. However, in \cite{lenoci2006self}, phase separation is not coupled with chemical reaction and diffusion, therefore there is a complete coarsening of organic-rich phase regions \cite{glotzer1995reaction}. In our study, we integrate the prepatterning field into the reaction-diffusion model coupled with phase separation to mimic the presence of costae, contributing to pore formation. Our results show that diatom valve pore structures can be qualitatively mimicked by a liquid-liquid phase separation model combined with a chemical reaction. Although accurately reproducing the observed diatom structures would necessitate a more complex model, our research indicates that features such as costae and pores can be produced in silico as a negative imprint of the phase-separated organic template.

Although our computer simulations provided valuable insights, it is important to recognise their limitations. A more sophisticated modelling approach is needed, particularly to address the complexities of nonlinear diffusion and phase separation dynamics, as well as to account for the electrostatic interactions between organic droplets and silica ribs and the growth of SDVs during valve development. As the diameter of the SDVs increases, the silica ribs grow radially, necessitating an extension of our model to include the growing domain, which has an interesting effect on pattern formation (see e.g., \cite{crampin1999reaction,crampin2002pattern,krause2019influence}). This should also include a recent study on branching morphogenesis \cite{babenko2024mechanism} and costae growth. Furthermore, comparing our work with mass-conserving reaction-diffusion systems (e.g., \cite{weyer2023coarsening}) holds considerable interest and can provide valuable insights into the underlying mechanisms.

By integrating theoretical modelling with \textit{in vitro} experimental validation, we can enhance our understanding of these captivating biological phenomena and potentially apply them in various fields such as biomimetics and materials science. Experimental studies are crucial for the precise calibration of specific model parameters, whereas the insights gained from our simulations may provide a solid foundation for future \textit{in vitro} methodologies. These approaches can strategically adjust critical factors such as the pH level and the concentration of reacting constituents, thereby controlling the biosilica formation process.

Furthermore, our approach holds promise for studies aimed at creating artificial SDV under \textit{in vitro} conditions.  The modelling performed within this framework can further facilitate the modelling of silica structures, in which the arrangement of organic aggregates acts as a template for silica formation following the addition of monosilicic acid to the system.

Our investigation into the role of phosphate ions and organic molecules in the formation of the diatom silica structure provided valuable insights into the mechanisms driving pattern formation. By exploring the hypothesis of organic template-driven silica morphogenesis, our study focused on the aggregation of organic molecules through a phase separation model, supported by \textit{in vitro} experiments highlighting the catalytic role of organic components.

The approach used in this study couples the phase separation process with a chemical reaction, allowing the phase-separating component to dissociate into two constituents. Unlike previous studies \cite{bobeth2020continuum,lenoci2006self}, our work involved detailed analysis to elucidate the conditions necessary for pattern formation and to understand how model parameters and prepatterning field affect pattern morphology and transitions between different types of patterns. 

The study in \cite{glotzer1995reaction} shows that the chemical reaction suppresses spinodal decomposition, limiting the ongoing coarsening of the phase domains during phase separation. Building on the framework presented by Bobeth et al.\cite{bobeth2020continuum}, which couples phase separation with chemical reactions, the objective of this study was to explore the effects of model parameters on pattern formation. We provided valuable insights into the interplay between the initial concentrations and dissociation coefficients that govern phase separation. Through stability analysis, we established conditions for pattern formation and assessed how the initial concentrations and dissociation rate affect the dispersion relation. Our numerical simulations identified five distinct pattern types and emphasised the role of the initial concentrations and the reverse reaction rate in the formation of these patterns. Experimental evidence indicated that the type of soluble silica and biomolecules significantly influence silica morphology under different conditions \cite{bernecker2010tailored,lechner2015silaffins,sumper2006learning}. Consequently, our findings provide a robust framework for integrating these insights into more advanced models and \textit{in vitro} experiments designed to clarify these morphological variations in silica. In our work, we demonstrated that the pattern morphology depends on the initial concentrations of organic molecules and phosphate ions. We observed that transitioning between pattern types requires significant changes in the concentrations of both reacting constituents. Increasing the concentration of any reactant enlarges the region dominated by the organic-rich phase, facilitating the formation of diverse organic-rich networks, each exhibiting distinct characteristics that reflect their pattern types.

In \cite{lenoci2006self}, phase separation was induced by an additional local field due to the existing silica costae in the base layer, leading to the formation of various two-dimensional patterns that closely resembled the observed valve structures. However, in \cite{lenoci2006self}, phase separation is not coupled with chemical reaction and diffusion, therefore there is a complete coarsening of organic-rich phase regions \cite{glotzer1995reaction}. In our study, we integrate the prepatterning field into the reaction-diffusion model coupled with phase separation to mimic the presence of costae, contributing to pore formation. Our results show that diatom valve pore structures can be qualitatively mimicked by a liquid-liquid phase separation model combined with a chemical reaction. Although accurately reproducing the observed diatom structures would necessitate a more complex model, our research indicates that features such as costae and pores can be produced in silico as a negative imprint of the phase-separated organic template.

Although our computer simulations provided valuable insights, it is important to recognise their limitations. A more sophisticated modelling approach is needed, particularly to address the complexities of nonlinear diffusion and phase separation dynamics, as well as to account for the electrostatic interactions between organic droplets and silica ribs and the growth of SDVs during valve development. As the diameter of the SDVs increases, the silica ribs grow radially, necessitating an extension of our model to include the growing domain, which has an interesting effect on pattern formation (see e.g., \cite{crampin1999reaction,crampin2002pattern,krause2019influence}). This should also include a recent study on branching morphogenesis \cite{babenko2024mechanism} and costae growth. Furthermore, comparing our work with mass-conserving reaction-diffusion systems (e.g., \cite{weyer2023coarsening}) holds considerable interest and can provide valuable insights into the underlying mechanisms.

By integrating theoretical modelling with \textit{in vitro} experimental validation, we can enhance our understanding of these captivating biological phenomena and potentially apply them in various fields such as biomimetics and materials science. Experimental studies are crucial for the precise calibration of specific model parameters, whereas the insights gained from our simulations may provide a solid foundation for future \textit{in vitro} methodologies. These approaches can strategically adjust critical factors such as the pH level and the concentration of reacting constituents, thereby controlling the biosilica formation process.

Furthermore, our approach holds promise for studies aimed at creating artificial SDV under \textit{in vitro} conditions.  The modelling performed within this framework can further facilitate the modelling of silica structures, in which the arrangement of organic aggregates acts as a template for silica formation following the addition of monosilicic acid to the system.

\appendix
\section*{Appendixes}

\renewcommand{\thesection}{S\arabic{section}}
\renewcommand{\thefigure}{S\arabic{figure}}
\setcounter{figure}{0}
\begin{figure}[h]
	\begin{center} \
	\begin{minipage}[h]{1\linewidth} 
		{\includegraphics[scale=1]{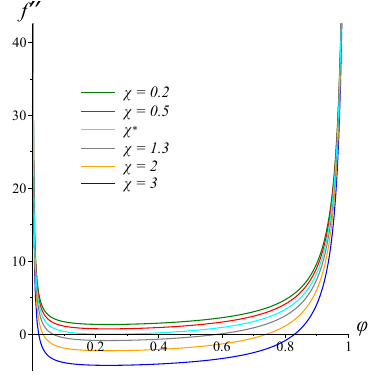}} 
	\end{minipage}
		\caption{Influence of Flory-Huggins interaction parameter on spinodal decomposition for $N=10$.\label{g(f)}}
	\end{center}
\end{figure}

\begin{figure*}[h]
	\begin{center} \
	\begin{minipage}[h]{1\linewidth} 
		{\includegraphics[scale=1]{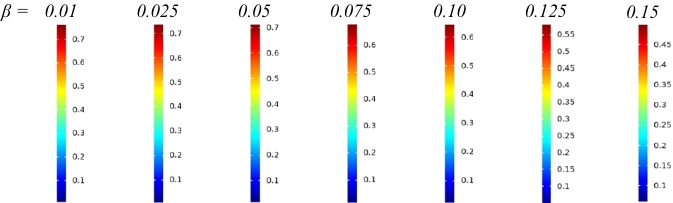}} 
	\end{minipage} 
		\caption{color bars: variation of concentration $\phi$ with  dissociation rate $\beta$} \label{color_bar}
	\end{center}
\end{figure*}

\section*{Numerical integration of PDEs}

To numerically integrate the system of nonlinear partial differential equations \eqref{ca_eq}-\eqref{phi_d}, we employed the CFD Module in Comsol Multiphysics. Using the backward Euler method in a nonlinear time-dependent solver, we sought to capture the dynamics of the system accurately. The boundary conditions and parameters used in the simulations are described in detail in the main text.

However, simulations of the initial system \eqref{ca_eq}-\eqref{phi_d} proved to be computationally intensive and slow, particularly at lower values of $\beta$. We adopted an approximate 4D system approach to accelerate the simulations without compromising accuracy. This approximated system, outlined below, yields solutions equivalent to those of the initial system, but with a significantly improved computational speed:
\begin{gather}
 \cfrac{\partial \widehat{c}_A }{\partial \widehat{t}}= \widehat{\nabla}  \cdot \widehat{D}_A \widehat{\nabla} \widehat{c}_A - \widehat{\alpha} \widehat{c}_A \widehat{c}_B + \widehat{\beta} \phi  \\ 
 \cfrac{\partial \widehat{c}_B }{\partial \widehat{t}}= \widehat{\nabla}  \cdot \widehat{D}_B \widehat{\nabla} \widehat{c}_B - \widehat{\alpha} \widehat{c}_A \widehat{c}_B + \widehat{\beta} \phi \\
 \cfrac{\partial \phi}{\partial \widehat{t}} = \widehat{\nabla}^2 \mu +  \widehat{\alpha} \widehat{c}_A \widehat{c}_B - \widehat{\beta} \phi\\
 \epsilon\cfrac{\partial \mu}{\partial \widehat{t}} +\mu =  \widehat{f}'(\phi) - \widehat{\nabla}^2 \phi ,
\end{gather}
where $\epsilon\ll 1$, in our calculations we put $\epsilon=10^{-4}$.
The solutions derived from the approximated 4D system are closely aligned with those of the original system, but the solver shows a significantly improved efficiency. This method enables an effective investigation of system dynamics and behaviour with reduced computational demands.

\bibliographystyle{abbrv} 
\bibliography{sample}

\end{document}